\documentclass[12pt]{article} \textheight 23cm
\usepackage{graphicx}
\usepackage{amsmath}
\usepackage{color}

\begin{document} \begin{center}
{\large \bf Topology of desiccation crack patterns in clay  and invariance of crack interface area with thickness
} \vskip 0.2cm
Tajkera Khatun$^1$, Tapati Dutta$^2$ and Sujata Tarafdar$^{1*}$\\
\vskip 0.2cm

$^1$ Condensed Matter Physics Research Centre, Physics Department, Jadavpur University, Kolkata 700032, India\\

$^2$ Physics Department, St. Xavier's College, Kolkata 700016, India\\

Tajkera Khatun: tajkerakhatun88@gmail.com,
Tapati Dutta: tapati$\_$mithu@yahoo.com\\
Sujata Tarafdar$^*$: Corresponding author, sujata$\_$tarafdar@hotmail.com, Phone 913324146666(Ex. 2760), Fax: 913324138917 

\end{center}
\vskip .2cm
\noindent {\bf Abstract}\\ 
 We study the crack patterns developed on desiccating films of an aqueous colloidal suspension of bentonite  on a glass substrate. Varying the thickness of the layer $h$ gives the following new and interesting results: (i)We identify a critical thickness $h_{c}$, above which isolated cracks join each other to form a fully connected network. A topological analysis of the crack network shows that the Euler number falls to a minimum at $h_{c}$. (ii) We find further, that the total vertical surface area of the clay $A_v$, which has opened up due to cracking, is a constant independent of the layer thickness for $h \geq h_c$. (iii) The total area of the glass substrate $A_s$, exposed by  the hierarchical sequence of cracks is also a constant for $h \geq h_c$. These results are shown to be consistent with a simple energy conservation argument, neglecting dissipative losses. (iv) Finally we show that if the crack pattern is viewed at successively finer resolution, the total  cumulative area of cracks visible at a certain resolution, scales with the layer thickness. A suspension of Laponite in methanol is found to exhibit similar salient features (i)-(iv), though in this case the crack initiation process for very thin layers is quite different.  
 	\section{Introduction}Formation of desiccation crack patterns is a well-studied and interesting subject which continues to receive considerable attention \cite{allen1986,book}. There are many factors which affect details of the crack patterns and kinetics of crack propagation, such as: composition of the desiccating material \cite{kitsunezaki,lopes,pauchard1,daniels,goehring}, temperature \cite{lee}, humidity, solvent, layer thickness \cite{groisman,allain,mal}, effect of field such as: mechanical \cite{naka1,naka2}, electrical \cite{taj1,taj2,taj3}, magnetic \cite{pauchard}, drying rate \cite{tarasevich} and substrate \cite{carle} among others. 	
 
It is well known that a suspension or slurry of clay or some granular material forms a pattern of cracks when left to dry.
 There are many studies investigating the effect of varying the thickness of the layer \cite{groisman,yow2010} and these have resulted in the following observations - first, that there is a critical cracking thickness $h_{cct}$ for a film , below which the film dries without cracking \cite{tirumkudulu}. Secondly, the average spacing between adjacent cracks is of the order of the film thickness. There are relatively few studies on how the \textit{crack widths} are affected on varying film thickness, though many workers note the hierarchical nature of the crack pattern where older cracks widen as new narrower cracks appear \cite{bohn,dibyendu}. Another common observation is that in a thicker layer, cracks are generally fewer in number but wider on the average \cite{ peel}.
 
 In this paper we report the effect of layer thickness on different features of desiccation crack patterns. Two systems have been studied - a slurry of bentonite clay in water and a suspension of Laponite in methanol. Both are allowed to dry on a glass substrate.
 
 For bentonite-water samples the thickness ($h$) of the dried layer varies between  0.295-0.890$\it{mm}$. There is a noticeable change in the crack patterns as  $h$ increases. There are no cracks when the layer is extremely thin. Above a critical cracking thickness $h_{cct}$  
  isolated cracks in the form of three-pronged stars begin to form. After a second critical thickness $h_c$, the isolated cracks become completely connected to form a closed network. As we increase the thickness further, the number of cracks decreases and they become wider. Analysis of the three-dimensional pattern shows that the total new vertical surface area ($A_v$) of the bentonite layer, exposed due to cracking  first increases with $h$ and then saturates to a constant value at $h=h_c$. In addition, the surface area ($A_s$) exposed on the glass substrate due to shrinking  of the clay is also a nearly constant quantity showing a behaviour similar to $A_v$. The result can be understood on the basis of a simple energy conservation argument, assuming fracture to be elastic. 
  
  After desiccation, when viewed from the top, the connected cracks form a hierarchical pattern, where cracks of varying width are present. Let us assume that the pattern is viewed at a certain 
resolution,  that is only cracks with widths above a certain minimum value $w_{min}$ are visible. We scale the total area covered by cracks $A_{cum}$ by $A_s$, and $w_{min}$ by the layer thickness $h$ for all $h \geq h_c$. If we plot scaled $A_{cum}$ versus the scaled $w_{min}$, all curves for different $h$ collapse to a single curve. This demonstrates the scale invariance of the crack pattern, which has been shown to be  fractal by various groups \cite{colina2000,baer2009,dibyendu}.

All of the above results are shown to hold true for another clay system, which is Laponite suspended in methanol. There is however a major difference between the two cases. For Laponite, $h_{cct}$ and $h_{c}$ could not be identified, here connected cracks persist up to the lowest thickness studied and finally the sample reduces to a fine powder on drying. 
We also evaluate the inter-facial energy of glass and bentonite-water colloidal solution by measuring the contact angle of the bentonite-water system on glass and the surface tension of the solution. These quantities are used in working out energy balance during crack formation.

In the first set of experiments, we use sodium-bentonite procured from Merck for the clay component.  
Bentonite is a geological term for soil material with a high content of swelling mineral which is usually montmorillonite \cite{vanolphan,olakarnland}. The general chemical formula of bentonite is $Na_{0.33}$[$Al_{1.67}Mg_{0.33}$]$Si_4O_{10}$ $[OH]_2$. In addition to montmorillonite, bentonite contains a small portion of other mineral matter, usually feldspar, volcanic glass, quartz, organic matter, gypsum etc. As the basic part of bentonite is montmorillonite, a description of montmorillonite is well applicable to bentonite. Montmorillonite is hydrous aluminium silicate containing small amount of alkali and alkaline material \cite{arthurclem}. A single unit cell of montmorillonite consists of one alumina sheet sandwiched between two silica sheets and that is why montmorillonite is called a 2:1 layer clay. Montmorillonite (hence bentonite) particles behave like negatively charged particles in the presence of water. The dispersed platelet size of montmorillonite is $\sim$ 0.8$\times$0.8$\times$0.001 microns.

The clay sample for the second set of experiments is a synthetic clay - Laponite (RD), from Rockwood Additives. Laponite has the chemical formula
$Na_{0.7}^+$ $[(Si_8Mg_{5.5}Li_{0.4})$ $O_{20}(OH)_4]^{0.7-}$ and a 2:1 layer structure very similar to montmorillonite. Laponite consists of nearly mono-disperse nano sized platelets of diameter $\sim 25 nm$ and thickness $\sim 1 nm$.
Details of the experiments and analysis are described in the following sections.
   \section{Materials and Methods}
 Our experiments consist of two sets - Set-I, where the desiccating material is a slurry of bentonite and water and Set-II, where we use Laponite and methanol. The complementary compositions Laponite in water and bentonite in methanol were not suitable. Laponite forms a gel in water, which shows hardly any cracks unless the sample is confined by bounding walls, or exposed to an electric field \cite{mal2007physa}, whereas for bentonite in water  random cracks form with average spacing depending on the film thickness $h$.
\subsection{Sample preparation}
\noindent {\textbf{Set-I: Bentonite in water}}\\
To prepare the bentonite water colloidal solution, $x$ g  of bentonite powder is added to $16x$ ml water  and it is allowed to soak for 2-3 hours and then stirred by a spatula. From the uniform colloidal solution we deposit various amounts on several petri dishes to get layers of different thickness. The glass petri dish has diameter 9 cm. During our experiments, the ambient temperature and humidity varied between 30 $^o$C - 32 $^o$C and 50 \% -60 \% respectively.\\ 
\noindent {\textbf{Set-II: Laponite in methanol}}\\
 The second set of experiments are done with a Laponite-methanol mixture. The procedure followed is exactly similar to that of bentonite described above. In place of bentonite and water we use Laponite and methanol respectively in the second set. Since Laponite readily forms a uniform suspension in methanol, there was no need to pre-soak the Laponite in methanol.
\subsection{Image analysis}
Photographs were taken by a digital camera Nikon COOLPIX L120 with 21X optical zoom. A 10X microscope was used for the measurement of layer thickness of the film.
The final crack patterns are analysed by ImageJ software. We grey scale the crack pattern choosing a proper threshold so that the cracks appear black while the peds are white. All our analysis is done on these grey-scaled pictures of the crack pattern.
 \section{Results}
 \begin{figure}[h]
\begin{center}
\includegraphics[width=12.0cm, angle=0]{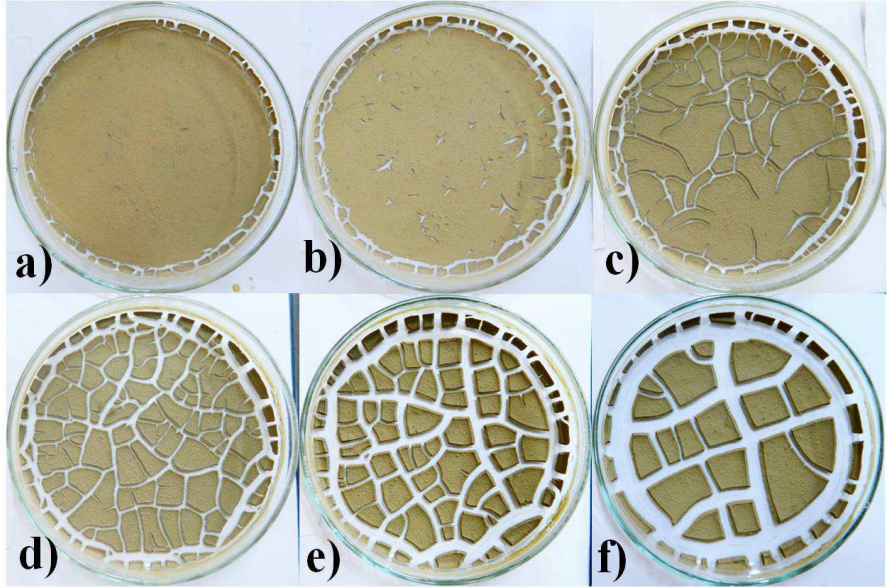}
\end{center}
\caption{Crack patterns for different layer thickness $h$ in bentonite-water (Set-I). $h$ = 0.0295$\it{cm}$ in a), 0.034$\it{cm}$ in b), 0.038$\it{cm}$ in c), 0.0405$\it{cm}$ in d), 0.0555$\it{cm}$ in e) and 0.089$\it{cm}$ in f) respectively.}
\label{bentcrk}
\end{figure}
 Results obtained for both Set-I and II are given below.
 The layer thickness ($h$) of the dried films in Set-I varies from approximately 0.295$\it{mm}$ to 0.890$\it{mm}$. For $h$ $\leq$ 0.295$\it{mm}$, no cracks appear except near the boundary of the petri-dish (Figure \ref{bentcrk}a). So the critical cracking thickness \cite{tirumkudulu,lewis,singh} for the bentonite water film may be taken as $h_{cct}=$ 0.295$\it{mm}$. 
 
\begin{figure}[h]
\begin{center}
\includegraphics[width=12.0cm, angle=0]{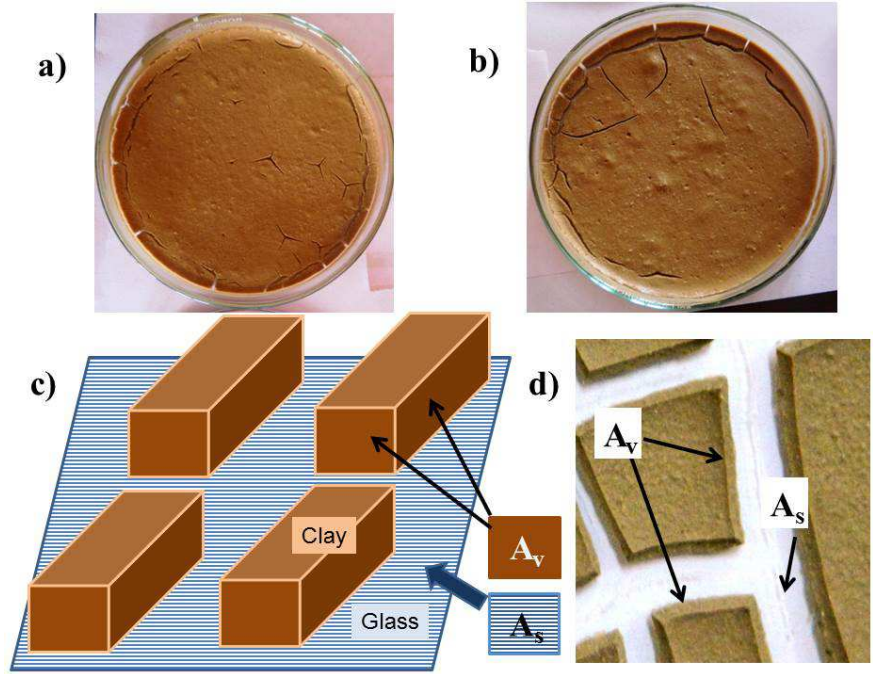}
\end{center}
\caption{Crack initiation in the sample, along with the magnified image of a small region for layer thickness $h$ a) $<$ $h_{c}$ and b) $>$ $h_{c}$ is shown for Set-I. c) Schematic diagram identifying crack area $A_v$ and $A_s$ and d) $A_v$ and $A_s$ marked on real crack patterns of the dried film of bentonite water slurry.}
\label{initiate}
\end{figure}

 As we increase $h$ above $h_{cct}$, isolated star-like cracks form first (Figure \ref{bentcrk}b) and they start to grow and connect among themselves (Figure \ref{bentcrk}c). This continues upto a certain value of $h$ which we call $h_{c}$. Beyond $h_{c}$, the  cracks finally form a closed network (Figure \ref{bentcrk}d). As $h$ increases beyond $h_{c}$, wider cracks form and the number of cracks decrease (Figure \ref{bentcrk}e-f). Crack initiation mechanisms are also different below and above $h_{c}$ as shown in Figure (\ref{initiate}a-b). Nearly straight random cracks first initiate for the film when $h$ is greater than $h_{c}$ whereas below $h_{c}$ star-like cracks form at the initiation.
 
In Set-II the layer thickness $h$ of the dried films varies approximately from 0.592$\it{mm}$ to 1.692$\it{mm}$. Here we cannot identify the critical cracking thickness ($h_{cct}$) nor the critical thickness ($h_c$). For all $h$ cracks form connected network. 
 
 The boundary wall of the petri-dish affects the crack pattern near the edges (Figure \ref{bentcrk}a-f). To eliminate this during image analysis, we discard an annular band near the periphery where the edge effect is prominent for both Set-I and II.
 We use ImageJ software to grey-scale the crack image using an appropriate threshold. Since the area eliminated to avoid the edge effect is not exactly equal for all samples, during quantitative analysis we divide all extensive quantities by the actual area considered. All results given henceforth for crack perimeter, total crack area, cumulative crack area etc. are normalized values which can be compared with each other meaningfully.\\
 \subsection{Euler number}
\begin{figure}[h]
\begin{center}
\includegraphics[width=12.0cm, angle=0]{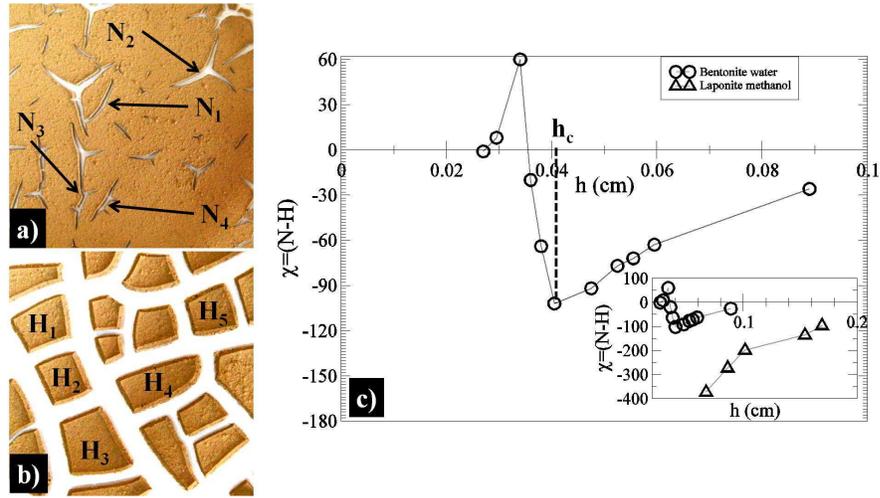}
\end{center}
\caption{a) Shows examples of isolated cracks $N_1$, $N_2$, $N_3$, $N_4$,... and b) shows peds $H_1$, $H_2$, $H_3$, $H_4$,.... on the film of bentonite-water colloidal solution (Set-I). c) Euler number $\chi$ vs. $h$ plot for the crack patterns of bentonite-water colloidal solution. Inset of c) represents the plot of $\chi$ vs. $h$ for the crack patterns of both Sets-I and II.}
\label{eulrn}
\end{figure}

 Vogel et al \cite{vogel} have shown that a topological description characterizing a crack network can be obtained using the Euler number defined as
   \begin{equation}
  \chi=N-H
  \end{equation} 
  where, $N$ is the total number of isolated cracks (Figure \ref{eulrn}a) and $H$ is the total number of peds i.e. solid blocks bounded by cracks (Figure \ref{eulrn}b).
  Figure (\ref{eulrn}c) shows the variation in $\chi$ with $h$ for Sets I and II (inset). The general behaviour can be described as follows. When there is no crack, i.e. for $h < h_{cct}$, $N=0$ and $H=1$, since the whole sample is a single ped. So initially $\chi=-1$. As isolated cracks appear, $N$ increases, but $H$ remains 1 as long as  the cracks do not join up to produce a new ped surrounded by cracks. This happens on increasing $h$ further. As the cracks join $N$ starts to decrease. As a result $\chi$ continues to fall. When all cracks join forming a fully connected network, the value of $N$ is 1. At this point, since $H$ has a positive value, $\chi$ reaches a minimum (i.e. most negative value). Further increase in $h$, however reduces $H$, as decrease in the density of cracks implies a decrease in the number of peds. It is to be noted that whatever the density of cracks, $N$ is to be counted as 1, as long as all cracks are connected. So the curve for $\chi$ climbs up again towards the X-axis. Figure (\ref{eulrn}c) shows exactly this behaviour for Set-I data. For Set-II data however, the cracks are always present and fully connected, so the graph for $\chi$ starts at a negative value, but climbs towards the X-axis, just like the Set-I data. For Set-I the minimum in $\chi$ can therefore be identified with a second critical thickness $h_{c}$, i.e. the threshold film thickness where the cracks first form a connected network.
  The threshold for a connected network of cracks is important in areas of soil science \cite{baer2009} and bioscience \cite{bio}. The Euler number $\chi$ may be useful in this context.  
\subsection{Perimeter $P_{cr}$ and Crack area $A_v$ and $A_s$} 
The sum total of the perimeter bounding all the peds is designated as the crack perimeter $P_{cr}$. As  cracks form, new interface area which forms the vertical walls of the clay peds  opens up (Figure \ref{initiate}c-d). We name the total vertical area of clay exposed for the sample as  $A_v$.
\begin{figure}[h]
\begin{center}
\includegraphics[width=10.0cm, angle=0]{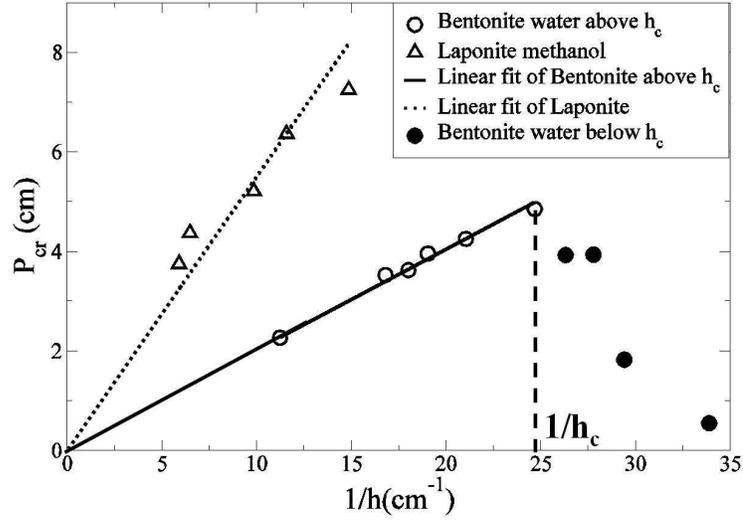}
\end{center}
\caption{$P_{cr}$ vs. $\frac{1}{h}$ plots for the crack patterns of Set-I and Set-II for all $h$. The solid and open circles represent the data for $h < h_c$ and $h \geq h_c$ respectively for Set-I. The solid and dotted lines represent the best fit straight lines for Set-I when $h \geq h_c$ and for Set-II when all $h$ respectively.}
\label{srfc-layer-crkprmtr}
\end{figure}
 The crack perimeter $P_{cr}$ is measured from the grey-scaled figures for every film thickness $h$, and  $A_v$ is calculated from 
\begin{equation}
 A_v=P_{cr} \times h
 \label{av}
 \end{equation} 
 For Set-I $P_{cr}$ increases with $h$ upto $h_c$ and then decreases . However, our experiments show that above $h_c$, though $P_{cr}$ falls with $h$, the product $P_{cr} \times h = A_v$ remains constant. The average $A_v$ for Sets I and II are respectively 0.20 and 0.57 cm$^2$ per unit sample area.
   $P_{cr}$ is plotted against $\frac{1}{h}$  in Figure (\ref{srfc-layer-crkprmtr}) for all $h$  for Sets I and  II. We see that the data for $h \geq h_c$ falls on a straight line passing through the origin, showing that equation (\ref{av}) is valid in both cases. The slopes of the two lines 0.2 and 0.55 match the values from the direct calculation.
  The straight line fit for Set-II is not as good as the Set-I fit. Possible reasons for this are discussed later in section (\ref{explanation}). 
 
\begin{figure}[h]
\begin{center}
\includegraphics[width=12.0cm, angle=0]{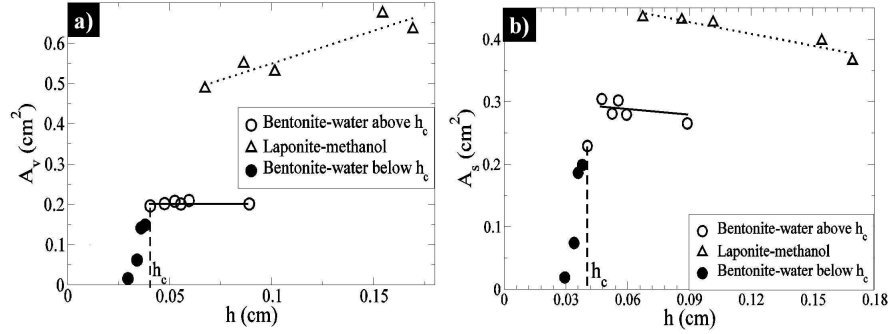}
\end{center}
\caption{a) and b) represent the variation of $A_v$ and $A_s$ with $h$ respectively for the crack patterns of Set-I and II for all $h$. The solid and open circles represent the data for $h < h_c$ and $h \geq h_c$ respectively for Set-I.}
\label{srfc-sbstrt}
\end{figure}

  As the clay film cracks, the underlying substrate surface is exposed.  The total substrate area exposed due to cracking, $A_s$ (Figure \ref{initiate}c-d), is measured directly from the  photographs using ImageJ. Figure (\ref{srfc-sbstrt}b) shows that  $A_s$ is also approximately constant for the range of $h \geq h_c$ studied for Set-I and for all $h$ of Set-II. $A_s$ behaves as an increasing function of $h$ below $h_c$. It is assumed that  $l_{cr}$ is the total length of the cracks measured along the center of the cracks. So the total crack length $l_{cr}$ can be taken as approximately half the value of $P_{cr}$ for each $h$. Then the average crack width is obtained from equation   
\begin{equation}
  A_s=l_{cr} \times w_{cr}=\frac{P_{cr}}{2} \times w_{cr}
  \label{as}
  \end{equation}  
From equations ({\ref{av}}) and (\ref{as}), we obtain
  \begin{equation}
  w_{cr}=\frac{2A_s}{A_v} \times h
  \label{wcr} 
 \end{equation}
 \begin{figure}[h]
\begin{center}
\includegraphics[width=10.0cm, angle=0]{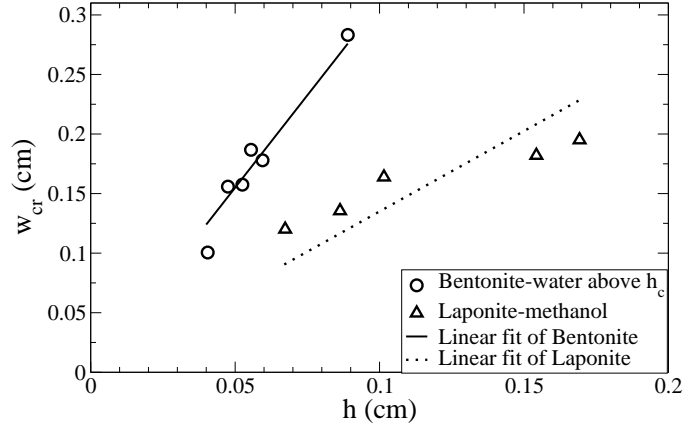}
\end{center}
\caption{Variation of average crack width $w_{cr}$ against $h$  for the crack patterns for Set-I when $h \geq h_c$ and for Set-II for all $h$. The solid and open circles represent the data for $h < h_c$ and $h \geq h_c$ respectively for Set-I. The solid and dotted lines represent best-fit lines for Set-I for $h \geq h_c$ and Set-II for all $h$ respectively.}
\label{wdth-thkns}
\end{figure}
  The variation of $w_{cr}$ with $h$ for the crack patterns of both Set-I (for $h \geq h_{c}$) and Set-II (for all $h$) are approximately linear as shown in Figure (\ref{wdth-thkns}). In case of Set-II the data shows large fluctuations about the best fit line through the origin. The possible explanation for these fluctuations is discussed in section \ref{explanation}.  Below $h_c$ for Set-I, the behaviour of $w_{cr}$ with $h$ is totally different. As $h \rightarrow 0$, $A_s$ and $A_v \rightarrow 0$, so $w_{cr}$ cannot be calculated (see equations \ref{as} and \ref{wcr}). 

 \subsection{Scaling of cumulative crack area}
\begin{figure}[h]
\begin{center}
\includegraphics[width=10.0cm, angle=0]{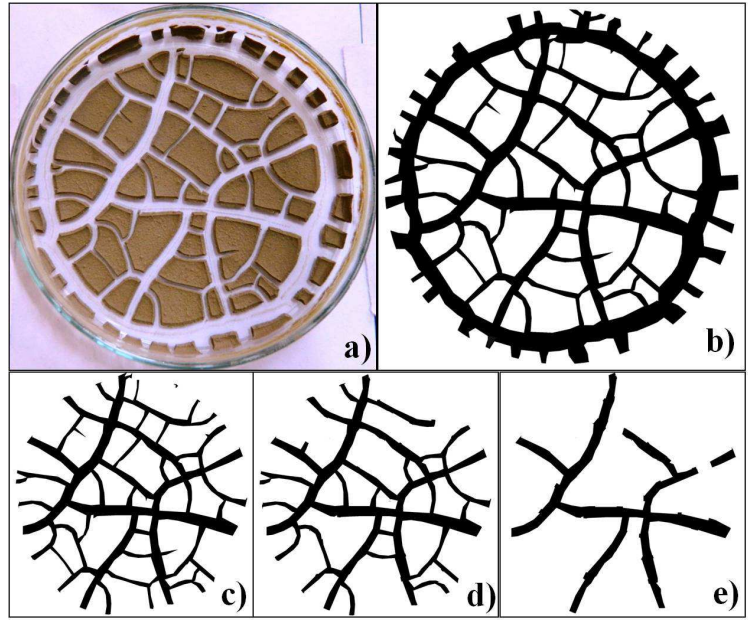}
\end{center}
\caption{Crack pattern for layer thickness $h$= 0.0595$\it{cm}$. a)-e) show the step by step procedure to measure cumulative area $A_{cum}$ by ImageJ software. b) represents the black and white photo of the crack pattern shown in a). The minimum crack width $w_{min}$ takes the following values,  for c) 0.0241$\it{cm}$}, d) 0.0995$\it{cm}$ and e) 0.1739$\it{cm}$.
\label{bentcrk-bw}
\end{figure}

\begin{figure}[h]
\begin{center}
\includegraphics[width=12.0cm, angle=0]{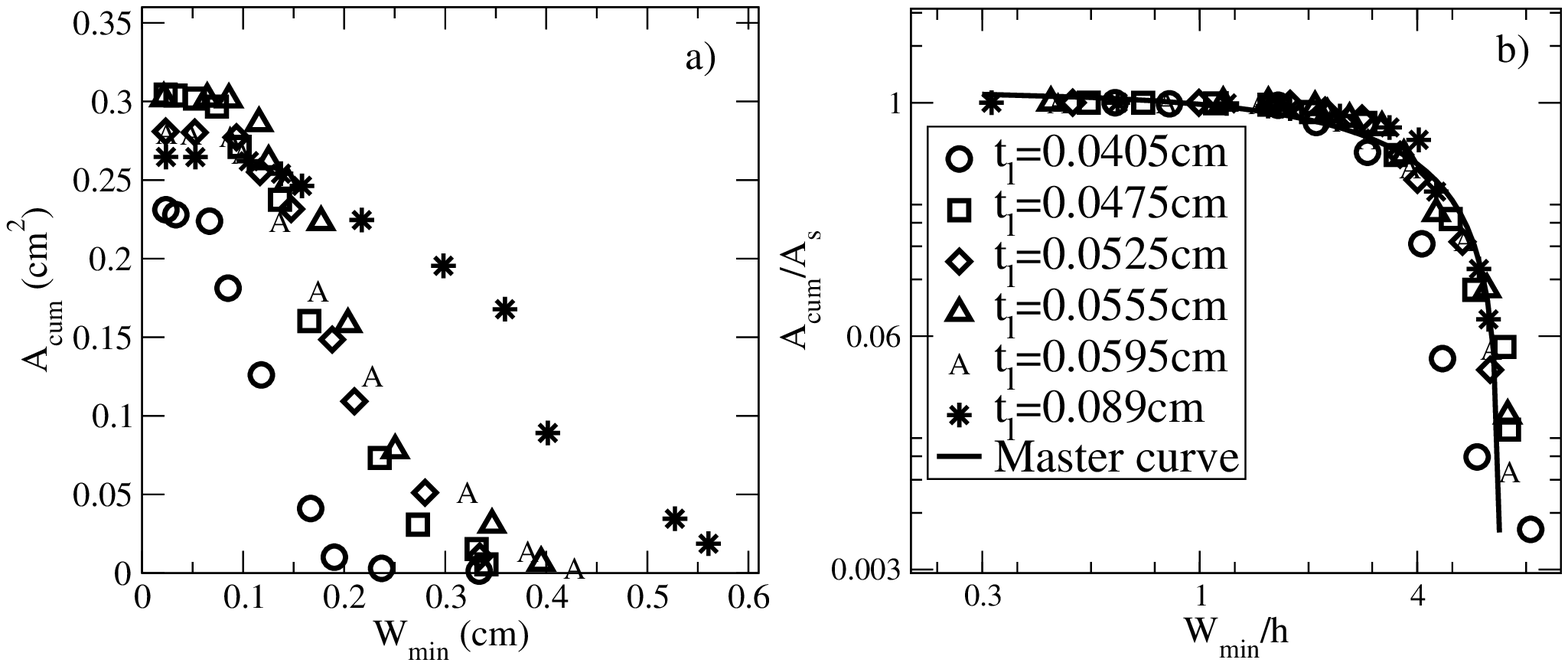}
\end{center}
\caption{a) Cumulative area of cracks ($A_{cum}$) plotted against minimum crack width $w_{min}$ for each layer thickness ($h \geq h_c$) from Set-I data. b) Plot when $A_{cum}$ is scaled by maximum cumulative area, which is equivalent to $A_{s}$ and $w_{min}$ is scaled by layer thickness ($h$). The solid curve in b) represents the master curve satisfying the relation $\frac{A_{cum}}{A_{s}}$ = 1.15 - 0.1696 $\frac{w_{min}}{h}$}
\label{cumulative}
\end{figure}

 The cumulative area ($A_{cum}$) of cracks with width above a certain minimum width ($w_{min}$) is the total area exposed on the substrate considering only cracks wider than $w_{min}$. Figure (\ref{bentcrk-bw}) shows the step by step procedure for measuring $A_{cum}$ for a certain $w_{min}$ at a particular $h$. The  behaviour of $A_{cum}$ with $w_{min}$ for various $h \geq h_c$ is shown in Figure (\ref{cumulative}a) for Set-I data. Here each curve represents a different value of $h \geq h_c$. 
  
\begin{figure}[h]
\begin{center}
\includegraphics[width=12.0cm, angle=0]{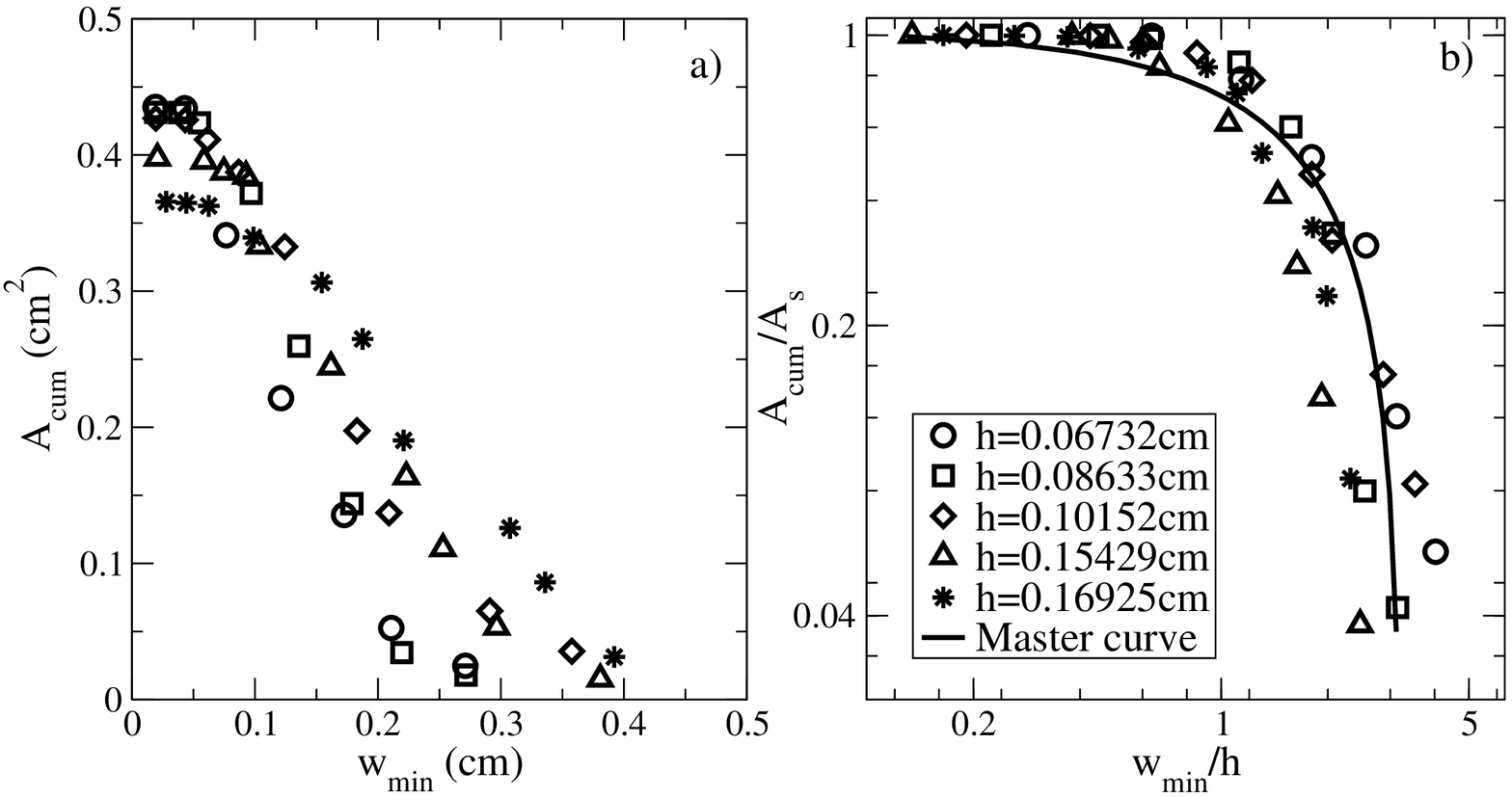}
\end{center}
\caption{a) Cumulative area of cracks ($A_{cum}$) plotted against minimum crack width $w_{min}$ for each layer thickness ($h$) from Set-II data. b) Plot when $A_{cum}$ is scaled by maximum cumulative area, which is equivalent to $A_{s}$ and $w_{min}$ is scaled by layer thickness ($h$). The solid curve in b) represents the master curve satisfying the relation $\frac{A_{cum}}{A_{s}}$ = 1.035 - 0.32 $\frac{w_{min}}{h}$}
\label{cumulative-area-lap}
\end{figure}

The interesting observation here is that all the curves for different $h$ merge together to an approximately single curve as shown in Figure (\ref{cumulative}b), when $A_{cum}$ is scaled by the maximum cumulative area which is nothing but $A_{s}$ and $w_{min}$ is scaled by $h$ in log-log scale. $A_{cum}$ is maximum when $w_{min}$ is the lowest crack width visible. On the other hand when only the widest crack is considered $A_{cum}$ is minimum.
  Similar behavior is obtained for Set-II data as shown in Figure (\ref{cumulative-area-lap}). 
  Similar observation was noted earlier by Mal et al \cite{dibyendu} for Laponite methanol mixtures.
  
The collapsed master curve can be fitted quite well by the linear equation,
\begin{equation}
\frac{A_{cum}}{A_s} = a\frac{w_{min}}{h} + b
\end{equation}

where, the constants $a$ = -0.1696 and -0.32 and $b$ = 1.15 and 1.035 for Set-I and Set-II respectively.  

The constant $b$ is the intercept of the above straight line on the Y-axis, i.e. the value of  $A_{cum}/A_s$ in the limit $w_{min}/h \rightarrow 0$. For a finite $h$, this limit means a vanishingly  small $w_{min}$, where all cracks are included. This implies $A_{cum}=A_s$ hence ideally $b$ should be 1. The best fit $b$ from Figures (\ref{cumulative}b) and (\ref{cumulative-area-lap}b) agrees approximately with this value. The non zero value of $b$ implies that even for very large $h$, we cannot get a fully crack free film. 
 \section{Evaluating energy spent and released in crack formation}
 We make an estimate of the energy released due to stress relaxation on cracking and the energy used up for creating new surface area during formation of the cracks. We assume that  the material is elastic.
 The net energy spent for the formation of cracks in both Sets I and II consists of two contributions: (1) energy to form the clay-air interface area $A_v$ and (2) energy required to expose the glass surface $A_s$. The net  energy spent ($U_{spent}$) necessary to create $A_v$ and $A_s$ is given by 
 \begin{equation}
 U_{spent}=G_{clay-air}A_v + (G_{gl-air}-G_{gl-clay})A_s
 \label{spent}
 \end{equation}
 where, the first term represents the energy needed to create $A_v$ and the second term is that to create $A_s$. $G_{clay-air}$, $G_{gl-air}$ and $G_{gl-clay}$ are the interface energies of the clay colloidal suspension in air, glass in air and glass in clay suspension respectively. 'Clay suspension' represents here bentonite in water for Set-I and Laponite in methanol for Set-II.
 
 We measure $G_{clay-air}$ by Kruss Tensiometer for Set-I and its value is $\sim$ 81.34 mJ/$m^2$. ($G_{gl-air}-G_{gl-clay}$) is calculated by measuring the contact angle ($\theta_c$) of a drop of bentonite water colloidal solution on a glass substrate, using the relation 
\begin{equation}
G_{gl-air}=G_{gl-clay} + G_{clay-air}cos{\theta_c}
\label{angle}
\end{equation}
The value of $\theta_c$ experimentally measured in our laboratory is $\sim$ 28$^o$ $\pm 1.5^o$ for bentonite-water. Substituting these values in equations (\ref{spent}) and (\ref{angle}), we get $U_{spent}\simeq 36.105$ mJ for Set-I for any $h \geq h_c$. 

According to Griffith's criterion \cite{tnsnrlse} crack formation is possible in elastic media when 
\begin{equation}
U_{released}\geq U_{spent}
\label{econserve}
\end{equation}
As a crack forms, it tends to release the stress in its vicinity \cite{tnsnrlse}. For a crack of length $l_{cr}$, width $w_{cr}$ and layer thickness $h$, it can be shown \cite{ereleased} that the total energy released $U_{released}$ is given by
\begin{equation}
U_{released}\sim\frac{{\sigma}^2l_{cr}w_{cr}h}{E}=\frac{{\sigma}^2A_sh}{E}=\frac{{\sigma}^2A_vw_{cr}}{2E}
\label{released}
\end{equation}
where, $E$ is the elastic coefficient of the material and $\sigma$ is the stress developed inside the material. In this case stress builds up due to desiccation. In our Set-I experiments  $A_s$ is constant for $h \geq h_c$, so $U_{released}$ depends on $h$ (or $w_{cr}$) and $\sigma$ whereas $U_{spent}$ is independent of $h$. So, if the energy balance equation is to hold $\sigma \propto 1/\sqrt{h}$. As desiccation proceeds, the stress builds up until it reaches the critical stress for fracture, so $\sigma$ in equation (\ref{released}) is this critical stress.

Earlier literature shows that $\sigma \propto 1/{h}^m$, where the value of the exponent $m$ is given as $1/2$ in \cite{evans}, assuming the material and substrate to have similar elastic properties. When the film and the substrate are purely elastic but with dissimilar properties \cite{tirumkudulu,singh}, $m$ takes a value $2/3$ which is not too different from $1/2$. We find therefore that our experimental results can be approximately explained assuming linear elastic fracture mechanics. More significantly, the constancy of the two surfaces created by fracture, irrespective of the layer thickness is consistent with this formalism.

For the Laponite methanol mixture, i.e. Set-II, there is a problem in following the same procedure. Due to rapid precipitation of the Laponite particles in the Laponite-methanol mixture,  we can neither measure $G_{clay-air}$ by the surface tensiometer nor can a droplet be deposited on the substrate to measure the contact angle $\theta_c$. Hence we cannot calculate $U_{spent}$ for Set-II. However the areas $A_v$ and $A_s$ are approximately constant in this case also, so the principal conclusion is likely to hold here as well.
\section{Discussion}
\label{explanation}
   In assessing the significance of this study, we may note that
  though there are a large number of studies of crack pattern variation with layer thickness \cite{allain,mal} reporting significant findings, several new and interesting observations have resulted from the present set of experiments. We have not come across measurements of the crack area $A_v$ or $A_s$ and the demonstration of their invariance with $h \geq h_c$ seems quite remarkable.
  
  The identification of the second critical thickness $h_c$ for Set-I, where the crack network becomes fully connected is a significant finding. The fact that $h_c$ or $h_{cct}$ were not observed for Set-II samples may be due to the difference of nearly three orders of magnitude in the particle sizes of Laponite and bentonite. It has been reported that $h_{cct}$ is of the order of the largest inhomogeneity in the sample \cite{inhomo}. In this case this may be taken as the average particle size, which is $\sim \mu m$ for bentonite and $\sim nm$ for Laponite.
  
  It is to be noted that the results for Set-II are less reliable compared to Set-I. This is because  Laponite and methanol do not mix and the drying sample remains inhomogeneous. This causes grains to separate out sometimes and
  in our experiments the area measured as $A_s$ is not completely clean. Some material remains on the glass surface in the middle of cracks. This is ignored in our analysis and the crack surfaces on the glass are assumed as clear. This leads to  some error in measured results. For Set-I, bentonite forms a uniform slurry and this problem is nearly absent. So it is expected that there is more error in the measured results of the crack patterns for Laponite-methanol mixture. This residual material makes a difference in the estimation of layer thickness and may be responsible for the slightly larger deviation in Set-II results compared to Set-I. It may be noted that the data for bentonite show a uniformly consistent behavior except the point for $h=0.405mm$ which is on the border line of the transition across $h=h_c$. This is noticeable particularly in Figures (\ref{srfc-sbstrt}) and (\ref{wdth-thkns}).
  
 Another interesting  result is the identification of the second critical thickness $h_c$, where all cracks connect. This is similar to a percolation transition \cite{stauffer}, with the connected crack network playing the role of an infinite cluster. Relating this transition point to the minimum value of the Euler number $\chi$ is also significant. It illustrates the role of topological concepts in crack networks and was initially proposed by Vogel et al.\cite{vogel}.
 
 Crack width ($w_{cr}$) varies linearly with $h$ beyond $h_{c}$ for cracks on the film of bentonite water colloidal solution, and for all $h$, for cracks on the film of Laponite methanol mixture.
 
 The scaling relation for the cumulative crack area measured at different resolutions with $h$ was reported earlier by Mal et al. \cite{dibyendu} for Laponite-methanol mixtures. Here it is demonstrated for bentonite-water slurries as well and shown to be reproducible for Laponite-methanol mixtures. The master curve obtained on collapsing the data can be fit to a linear relation between the scaled cumulative area ($\frac{A_{cum}}{A_s}$) and $\frac{w_{min}}{h}$, here $h$ is always $\geq h_c$ . This may be correlated to the scale invariance of the pattern as expected for a fractal system. Experiments on different substrates may lead to further significant findings.
 
 Assuming the system to be elastic, 
 the energy spent $U_{spent}$ for crack formation is constant for a given clay-solvent combination and can be determined from the experimental measurements done on crack patterns. Setting up the energy balance relation, dimensional arguments show that invariance of the area $A_v$ and $A_s$  agrees with earlier estimates of the dependence of $\sigma$ as a function of $h$.
 
 The above analysis of the results, invoking the energy inequality equation (\ref{econserve}), may face the criticism that energy dissipation has been totally ignored. It has been shown \cite{kitsunezaki,so} that in such situations plastic deformation and other forms of energy dissipation may be quite important.
 
 To look for one possible signature of energy dissipation, 
  micro-graphs of randomly selected regions of the film have been acquired using FESEM (Configuration No. QUO-35357-0614 funded by FIST-2, DST) to assess the effect of the layer thickness on micro-structure, if any. However, no noticeable changes could be discerned in the structure or density of micro-cracks, which may have been responsible for additional energy dissipation. The micro-graphs of films of different thickness look very similar.
  \section{Conclusions}
  We conclude by summarizing the salient points of our study. We studied crack patterns generated by desiccation in two systems on a glass substrate - Set-I: a suspension of bentonite in water and Set-II: a mixture of Laponite and methanol. The thickness of the drying layer $h$ was varied upto $\sim 2 mm$.
 \begin{enumerate}
 {\item For Set-I two critical values of $h$ were identified - $h_{cct}$, below which there are no cracks and $h_c>h_{cct}$, where the cracks first form a connected network. Here the Euler number for the network of cracks has a minimum at $h_c$. For Set-II connected cracks are observed down to the lowest $h$ studied. }
 {\item The new interface area generated by the cracks, $A_v$ for clay-clay debonding and $A_s$ for clay-glass debonding are independent of $h$ for connected network of cracks. This is shown to be consistent with earlier work on the dependence of critical cracking stress on $h$.}
   {\item The cumulative area of cracks $A_{cum}$ measured at a definite resolution, i.e. restricting measurement to a minimum observable thickness $w_{min}$, scales linearly with $h$. This is a manifestation of the scale invariance, i.e. fractal nature of the crack patterns.  }
It would be interesting to see if these observations hold for other desiccating systems.
\end{enumerate} 
  \section{Acknowledgement}
 TK thanks CSIR for providing a research grant. This work is supported by a joint Indo-Japan collaboration under DST-JSPS, authors are grateful to Akio Nakahara, So Kitsunezaki and Lucas Goehring for stimulating discussion. Authors thank T.R. Middya, Department of Physics, Jadavpur University, for support and useful suggestions. Micrograph images are taken using the FESEM  facility, configuration no. QUO-35357-0614 funded by FIST-2, DST Government of India, at the Physics Department, Jadavpur University.

\end{document}